\documentclass[draft]{agujournal2018}
\draftfalse
\usepackage{url}
\usepackage{lineno}
\usepackage{booktabs}
\usepackage{amssymb}

\usepackage{journal_macros}

\usepackage[bookmarks=false, pdfnewwindow=true, colorlinks=true, linkcolor=magenta, citecolor=blue, filecolor=cyan, urlcolor=purple, final=true]{hyperref}

\journalname{Space Weather}

\newcommand\editone[1]{\textcolor{black}{#1}}

\graphicspath{{./}{figures/}}

\begin{document}

\title{CME Evolution in the Structured Heliosphere and Effects at Earth and Mars During Solar Minimum}

\authors{Erika~Palmerio\affil{1,2,3}, Christina~O.~Lee\affil{2}, Ian~G.~Richardson\affil{4,5}, Teresa~Nieves-Chinchilla\affil{5}, Luiz~F.~G.~Dos~Santos\affil{6}, Jacob~R.~Gruesbeck\affil{7}, Nariaki~V.~Nitta\affil{8}, M.~Leila~Mays\affil{5}, Jasper~S.~Halekas\affil{9}, Cary~Zeitlin\affil{10}, Shaosui~Xu\affil{2}, Mats~Holmstr{\"o}m\affil{11}, Yoshifumi~Futaana\affil{11}, Tamitha~Mulligan\affil{12,13}, Benjamin~J.~Lynch\affil{2}, and Janet~G.~Luhmann\affil{2}}

\affiliation{1}{Predictive Science Inc., San Diego, CA, USA}
\affiliation{2}{Space Sciences Laboratory, University of California--Berkeley, Berkeley, CA, USA}
\affiliation{3}{CPAESS, University Corporation for Atmospheric Research, Boulder, CO, USA}
\affiliation{4}{Department of Astronomy, University of Maryland, College Park, MD, USA}
\affiliation{5}{Heliospheric Physics Division, NASA Goddard Space Flight Center, Greenbelt, MD, USA}
\affiliation{6}{nextSource Inc., New York, NY, USA}
\affiliation{7}{Solar System Exploration Division, NASA Goddard Space Flight Center, Greenbelt, MD, USA}
\affiliation{8}{Lockheed Martin Solar and Astrophysics Laboratory, Palo Alto, CA, USA}
\affiliation{9}{Department of Physics and Astronomy, University of Iowa, Iowa City, IA, USA}
\affiliation{10}{Leidos Innovations Corp., Houston, TX, USA}
\affiliation{11}{Swedish Institute of Space Physics, Kiruna, Sweden}
\affiliation{12}{Space Sciences Department, The Aerospace Corporation, Los Angeles, CA, USA}
\affiliation{13}{Department of Earth Sciences, Millersville University, Millersville, PA, USA}

\correspondingauthor{E.~Palmerio}{epalmerio@predsci.com}


\begin{keypoints}
\item We analyse the eruption and propagation of two CMEs from the Sun up to Earth and Mars during August 2018
\item Both CMEs were observed at Earth, but the second largely missed Mars, possibly due to interaction with a following high-speed solar wind stream
\item The sequence of events observed resulted in a strong magnetic storm at Earth, but only moderate disturbances at Mars
\end{keypoints}


\begin{abstract}
The activity of the Sun alternates between a solar minimum and a solar maximum, the former corresponding to a period of ``quieter'' status of the heliosphere. During solar minimum, it is in principle more straightforward to follow eruptive events and solar wind structures from their birth at the Sun throughout their interplanetary journey. In this paper, we report analysis of the origin, evolution, and heliospheric impact of a series of solar transient events that took place during the second half of August 2018, i.e.\ in the midst of the late declining phase of Solar~Cycle~24. In particular, we focus on two successive coronal mass ejections (CMEs) and a following high-speed stream (HSS) on their way towards Earth and Mars. We find that the first CME impacted both planets, whilst the second caused a strong magnetic storm at Earth and went on to miss Mars, which nevertheless experienced space weather effects from the stream interacting region (SIR) preceding the HSS. Analysis of remote-sensing and in-situ data supported by heliospheric modelling suggests that CME--HSS interaction resulted in the second CME rotating and deflecting in interplanetary space, highlighting that \editone{accurately reproducing} the ambient solar wind is crucial even during ``simpler'' solar minimum periods. Lastly, we discuss the upstream solar wind conditions and transient structures responsible for driving space weather effects at Earth and Mars.
\end{abstract}


\begin{plainlanguagesummary}
The Sun is characterised by a 11-year periodicity of its levels of activity, resulting in a solar minimum and a solar maximum alternating approximately every 5.5 years. During solar minimum, the Sun and its whole environment are in their simplest configuration, and eruptive events are significantly less frequent. It follows that periods of lower activity are generally considered optimal for tracking solar phenomena from their origin at the Sun throughout their journey in interplanetary space. In this paper, we analyse a series of solar eruptions that took place during the second half of August 2018 and follow them until their arrival at Earth and Mars, taking into account their associated effects on the two planets. We find that, even during solar minimum, the large-scale structure of the solar and interplanetary environment can have more or less dramatic impacts on the evolution of eruptions as they travel away from the Sun. Additionally, we suggest that the same event can cause diverse levels of disturbances at different \editone{planets}, depending on the particular structure and properties of the impacting solar wind.
\end{plainlanguagesummary}


\section{Introduction} \label{sec:intro}

The solar activity is characterised by numerous short- and long-term periodicities, the most renowed of which is the 11-year solar magnetic activity cycle \citep{hathaway2015}. Over the duration of a full cycle, \editone{the number of sunspots as well as the fraction of solar surface covered by them} rise until reaching a maximum and then fall again. This trend is also followed by the occurrence of solar eruptions, including flares \citep[e.g.,][]{benz2017} and coronal mass ejections \citep[CMES; e.g.,][]{webb2012}, which tend to peak close to solar maximum and decrease drastically around solar minimum. Due to the general lack of active regions, CMEs during solar minimum tend to be of the slow, streamer-blowout kind \citep[e.g.,][]{vourlidas2018}, although some major active-region eruptions may still occur \citep[e.g.,][]{nitta2011}. Solar minimum periods are characterised by a simpler configuration of the solar magnetic field, a generally slower and less variable solar wind, and a less energised space environment \citep{riley2001, riley2022}. These aspects make minima excellent times for tracing solar phenomena ``from start to finish'' and for defining the baseline heliophysical system, giving rise to large, coordinated initiatives such as the Whole Sun Month \citep[WSM;][]{galvin1999}, which took place during the cycle 22--23 minimum, and the Whole Heliosphere Interval \citep[WHI;][]{thompson2011}, which took place during the cycle 23--24 minimum. CME occurrence usually drops to about one per week, meaning that CME--CME interactions tend to be significantly less likely and that it is possible to follow the evolution of single CMEs and their interaction with solar wind structures, including the slow wind, high-speed streams (HSSs), and slow--fast stream interaction regions \citep[SIRs; e.g.,][]{richardson2018}.

During the cycle 24--25 minimum, \editone{the availability of space- and ground-based assets at various planets or scattered throughout the heliosphere, together with the growing consensus about the importance of multi-point studies within the heliosphere as a whole}, have led to the establishment of an even more comprehensive follow-up effort to WSM and WHI, i.e.\ the Whole Heliosphere and Planetary Interactions (WHPI; \url{https://whpi.hao.ucar.edu/}) initiative, which aims to study the interconnected solar--heliospheric--planetary system. In particular, a relatively large fleet of instruments was and is still operational at Mars, providing us with the opportunity to follow solar transients from the Sun to Earth and/or Mars and to characterise their space weather response at the two planets. Occasionally, the same transient may even encounter both Earth and Mars and elicit dissimilar impacts at each planet. Earth is characterised by a strong intrinsic quasi-dipolar magnetic field that is able to sustain a full-fledged magnetosphere \citep[e.g.,][]{pulkkinen2007}. Since Earth’s field is roughly directed towards the North in the \editone{equatorial} plane at the magnetopause, the most geoeffective solar wind structures are those containing southward magnetic field \citep{zhang2007} as well as high speed and ram pressure \citep{gonzalez1994}. Mars, on the other hand, lacks a global magnetic dipole and interaction between the solar wind and the Martian ionosphere generates a so-called induced magnetosphere \citep[e.g.,][]{bertucci2011}, which however differs from a Venus- or comet-like one due to the presence of localised crustal magnetic fields \citep{acuna1998}, thus leading to a ``hybrid'' magnetosphere \citep[e.g.,][]{dibraccio2018}. As a result, the main parameters that regulate the \textsl{arieffectiveness}---from the Greek name for Mars, ${'}\!\!A\rho\eta\varsigma$ or {\'A}ris---of a solar wind transient are the dynamic pressure \citep{opgenoorth2013}, usually enhanced in CMEs and SIRs, and the topology of the interplanetary magnetic field \citep{jakosky2015a}.

Despite the recent increasing interest in space weather at Mars \citep[e.g.,][]{geyer2021,huang2021,lee2017,luhmann2017,zhao2021}, most studies have focused on the analysis of the consequences of large and clear solar events \citep[e.g.,][]{crider2005,jakosky2015a,lee2018}, whilst less is known about minor events observed during periods of low solar activity \citep[e.g.,][]{kajdic2021,sanchezcano2017}. In order to tackle this issue, we present in this article a detailed analysis of a sequence of solar transients during the second half of August 2018 (i.e.\, during the late declining phase of Solar Cycle 24) between Earth and Mars, which were separated by ${\sim}8^{\circ}$ in longitude, ${\sim}2^{\circ}$ in latitude, and ${\sim}0.4$~AU in radial distance. Some aspects of these events were addressed by a number of studies \citep[e.g.,][]{abunin2020, akala2021, chen2019, cherniak2022, gopalswamy2022, mishra2019, moro2022, nitta2021, piersanti2020, thampi2021, younas2020, zhang2020}, and our goal here is to provide a holistic Sun-to-Mars investigation of their evolution from the Sun through the inner heliosphere and their effects at the two planets. Accordingly, we first present remote-sensing observations of the eruptive events. Then, we summarise the heliospheric context necessary to interpret in-situ observations at Earth and Mars. Then, we show solar wind and interplanetary magnetic field measurements at the two planets, with particular attention to the space weather responses to the interplanetary disturbances. Finally, we discuss these observations within the larger context of terrestrial and martian space weather during solar minimum periods.


\section{Remote-Sensing Observations} \label{sec:remotesensing}

The 20 August 2018 eruptive events were observed in remote-sensing imagery from two vantage points, namely Earth and the Solar Terrestrial Relations Observatory Ahead \citep[STEREO-A;][]{kaiser2008} spacecraft, located ${\sim}110^{\circ}$ east of Earth close to 1~AU. Here, we provide an overview of the sequence of events from their origin at the Sun (Section~\ref{subsec:disc}) through their evolution across the solar corona (Section~\ref{subsec:corona}) and inner heliosphere (Section~\ref{subsec:helio}).

\subsection{Solar Disc} \label{subsec:disc}

\begin{figure}[ht]
\centering
\includegraphics[width=.99\linewidth]{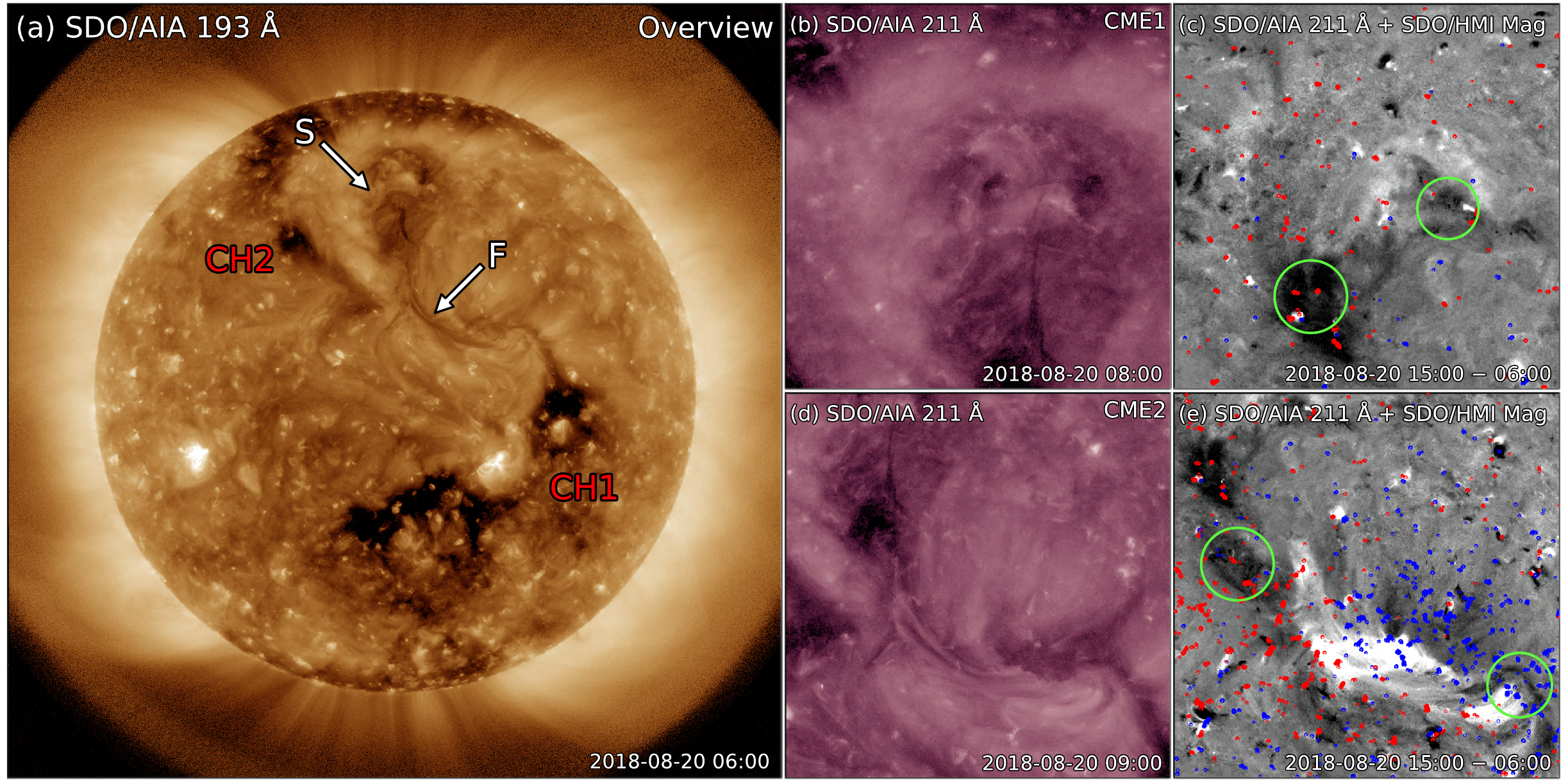}
\caption{Overview of the 2018~August~20 eruptions from remote-sensing solar disc imagery. (a) Pre-eruptive configuration on the solar disc. The erupting S-shaped structure (`S') and filament (`F'), as well as the two nearby coronal holes (`CH1' and `CH2') are labelled. (b) and (d) Zoomed-in images of the erupting sigmoid (CME1) and filament (CME2), respectively. (c) and (e) Base-difference images of the sigmoid and filament eruptions, respectively, with magnetogram data \editone{saturated to ${\pm}100$~G} overlaid (red = positive, blue = negative). The approximate eruption footpoints are indicated with green circles in both panels.}
\label{fig:sun}
\end{figure}

The sequence of eruptive events analysed in this study commenced on 20 August 2018 from an extended quiet-Sun region in the northern hemisphere located close to the central meridian from Earth's perspective. Figure~\ref{fig:sun}(a) shows the pre-eruptive configuration in extreme ultra-violet (EUV) at 193~{\AA} as seen by the Atmospheric Imaging Assembly \citep[AIA;][]{lemen2012} onboard the Solar Dynamics Observatory \citep[SDO;][]{pesnell2012} in orbit around Earth. The most striking features that were present on the Earth-facing Sun are an extended filament channel (marked as `F') and two coronal holes (CHs), a large one to the southwest of the filament (marked as `CH1') and a smaller one to the northeast (marked as `CH2'). Additionally, we note an S-shaped feature to the north of the filament (marked as `S'), reminiscent of a sigmoid \citep[e.g.,][]{green2007}---this structure is visible in several SDO/AIA channels, including 94~{\AA} and 131~{\AA}, which are known to respond to hotter plasma \citep{odwyer2010}. 

Both structures (i.e., the S-shaped feature and the filament) began erupting essentially at the same time, around 07:30~UT on 20 August, possibly in a sympathetic fashion \citep[e.g.,][]{torok2011,lynch2013}. Whilst the S-shaped structure (Figure~\ref{fig:sun}(b)) left the Sun rather rapidly (no signatures of the event were visible after a few hours), the filament (Figure~\ref{fig:sun}(d)) was involved in a much slower eruption, with post-eruption arcades still developing many hours after the onset and into the following day. Therefore, in this work we will define the S-shaped feature eruption as CME1, and the filament eruption as CME2. Around 18:30~UT on 20 August, a small portion of filament lying between the source regions of CME1 and CME2 erupted in a jet-like fashion \citep[see][for details]{mishra2019}---this may be considered as a ``second part'' of a two-step filament eruption, as suggested by \citet{abunin2020}. The full sequence of events observed by SDO is shown in EUV at 211~{\AA} in Movie S1.

Analysis of the pre-eruptive magnetic configuration of the two CMEs \citep[see][and references therein]{palmerio2017} suggests that CME1 was characterised by right-handed chirality (note its forward-S shape), whilst CME2 was left-handed (note its reverse S-shape). The approximate eruption footpoints \citep[estimated via the locations of coronal dimmings; e.g.,][]{reinard2009,thompson2000} and the magnetic field polarities in which they are rooted are shown in Figure~\ref{fig:sun}(c) for CME1 and Figure~\ref{fig:sun}(e) for CME2. Given that CME2 originated from a decayed active region, it is evident that its northeastern (southwestern) footpoint is rooted in a patch of positive (negative) polarity. Thus, its flux rope magnetic configuration upon eruption \editone{should be} west--south--east (WSE), following the convention of \citet{bothmer1998} and \citet{mulligan1998}. CME1, on the other hand, originated from a region of much weaker, quiet-Sun magnetic field, making determination of its magnetic structure and orientation more difficult. Nevertheless, the easternmost footpoint seems to be rooted in a patch of positive polarity, which for a right-handed flux rope would result somewhere between a west--north--east (WNE) and a south--west--north (SWN) type.

\subsection{Solar Corona} \label{subsec:corona}

\begin{figure}[ht]
\centering
\includegraphics[width=.99\linewidth]{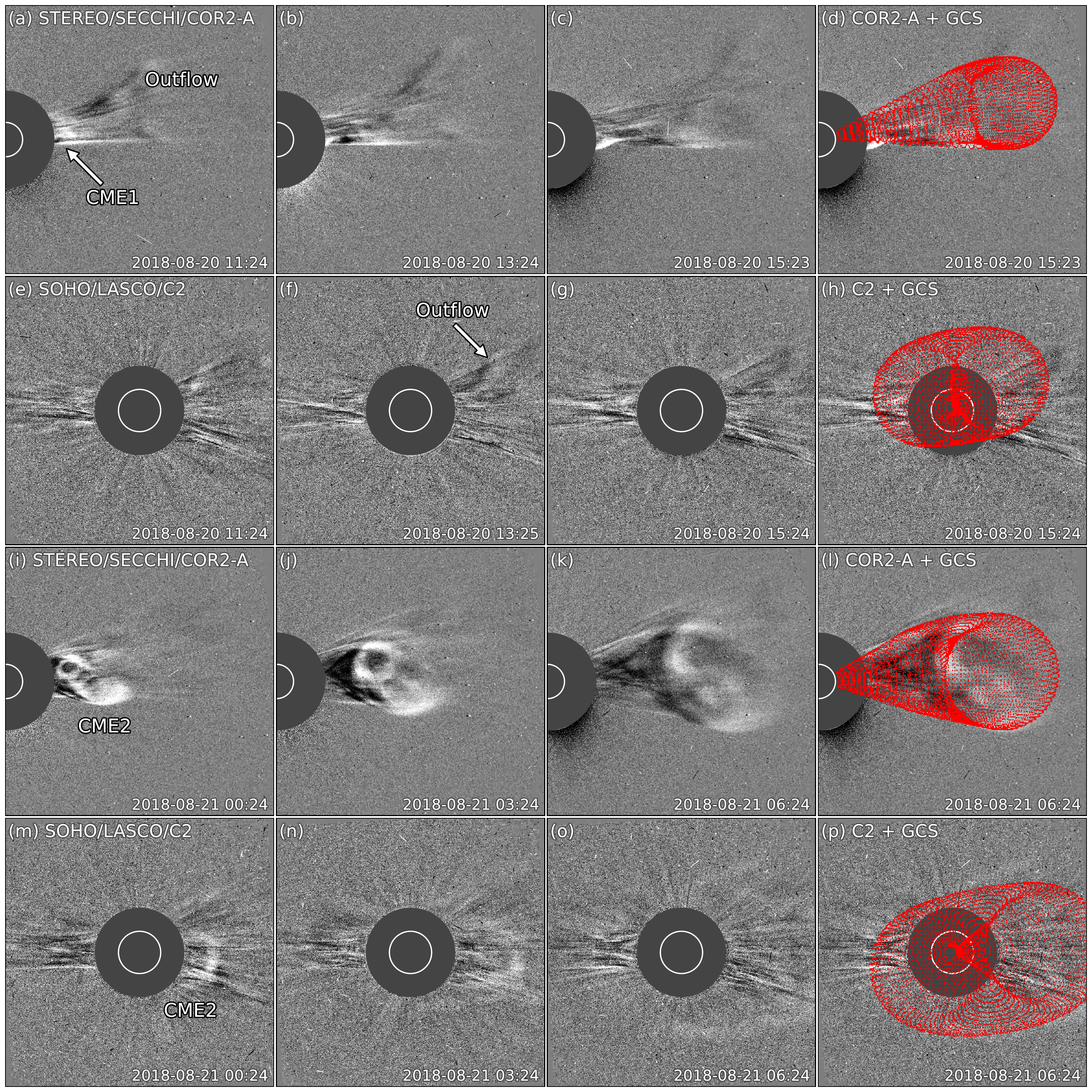}
\caption{Overview of the 2018~August~20 eruptions from remote-sensing coronagraph imagery (shown here in running difference with $\Delta t = 1$~hr). The first appearances of the preceding outflow, CME1, and CME2 in the field of view each observing instrument are indicated in the respective panels. The rightmost column shows coronagraph images with the Graduated Cylindrical Shell wireframes overlaid, showing the reconstruction of CME1 in panels (d) and (h) and CME2 in panels (l) and (p).}
\label{fig:corona}
\end{figure}

At the time of the 20 August 2018 eruptive events, white-light imagery of the solar corona was available from the Large Angle and Spectrometric Coronagraph \citep[LASCO;][]{brueckner1995} onboard the Solar and Heliospheric Observatory \citep[SOHO;][]{domingo1995}, located at Earth's Lagrange L1 point, as well as the coronagraphs forming part of the Sun Earth Connection Coronal and Heliospheric Investigation \citep[SECCHI;][]{howard2008a} suite onboard STEREO-A. Given that both CME1 and CME2 originated close to the central meridian as seen from Earth, we expect SOHO to have observed the eruptions approximately along their propagation direction, whilst STEREO-A had a near-quadrature view of the events. Such observations are summarised in Figure~\ref{fig:corona}, and the full sequence of imagery from STEREO/SECCHI/COR2-A and SOHO/LASCO/C2 is provided in Movie~S2 and Movie~S3, respectively.

First of all, we note the presence of coronal outflows preceding the eruption(s), indicated in Figure~\ref{fig:corona}(a,f) and possibly associated with blobs originating from the cusp of the helmet streamer belt \citep[e.g.,][]{lynch2020,wang2000}. CME1 first appeared in the COR2-A field of view around 10:24~UT on 20 August (Figure~\ref{fig:corona}(a--c)), featuring a rather irregular morphology \citep[it could be classified as a `jet' according to the definition of][]{vourlidas2013,vourlidas2017}. Furthermore, it appeared very faintly in STEREO-A imagery (and is more visible in Movie~S2 than in the still images shown in Figure~\ref{fig:corona}(a--c)), likely partially due to the presence of the preceding outflow along the line of sight, and was not clearly discernible in SOHO data (Figure~\ref{fig:corona}(e--g)). We note that CME events in the solar corona that appear faint/jet-like from one viewpoint and are not visible at all in another have been reported in previous studies, and may cause moderate geomagnetic disturbances \citep[e.g.,][]{palmerio2019}.

CME2 first appeared in the COR2-A field of view around 13:24~UT on 20 August (Figure~\ref{fig:corona}(i--k)) and in the C2 field of view around 21:48~UT on the same day (Figure~\ref{fig:corona}(m--o)). This event was significantly more evident than CME1 in images from both perspectives, with a flux rope-like morphology observed by STEREO-A and a full halo (albeit very faint) detected by SOHO. In particular, COR2-A imagery reveals an initially asymmetric structure, with its southern leg ahead of the northern one, that slowly swells before finally accelerating away from the Sun in a streamer-blowout fashion \citep[e.g.,][]{vourlidas2018}. As the CME travelled through the solar corona, its front became progressively less asymmetric, possibly indicating that either the northern leg caught up with the southern one, or that the whole structure rotated during its early propagation.

In order to estimate the geometric and kinematic properties of CME1 and CME2 through the solar corona, we fit both eruptions as they appeared in coronagraph imagery with the Graduated Cylindrical Shell \citep[GCS;][]{thernisien2006,thernisien2009} model. This will also serve to determine the CME input parameters needed for heliospheric modelling of their propagation through the structured solar wind (see Section~\ref{sec:enlil}). The GCS model consists of a parametrised shell (with six free parameters) intended to reproduce the flux rope morphology of CMEs, which can be applied to one or more nearly-simultaneous images and visually adjusted until its projection onto each field of view best matches the observations. Examples of GCS results are shown in the rightmost column of Figure~\ref{fig:corona}, for both CME1 (Figure~\ref{fig:corona}(d,h)) and CME2 (Figure~\ref{fig:corona}(l,p)). Both eruptions are estimated to have a low inclination to the solar equatorial plane in the outer solar corona, and the CME2 results indicate that it is larger (extending ${\sim}60^{\circ}$ versus ${\sim}40^{\circ}$ along the axis) but slower (${\sim}300$~km$\cdot$s$^{-1}$ versus ${\sim}500$~km$\cdot$s$^{-1}$) than CME1. We remark, however, that the GCS model is applied to single-point (STEREO-A) measurements for CME1 (Figure~\ref{fig:corona}(h) simply shows the projection of the shell in Figure~\ref{fig:corona}(d) onto the C2 field of view), inevitably resulting in larger uncertainties. Considering the flux rope types estimated at the Sun (see Section~\ref{subsec:disc}) and the low inclination of both eruptions through the solar corona, CME1 is expected to have an SWN configuration, whilst CME2 would reach a north--west--south (NWS) orientation via a counterclockwise rotation of its axis \citep[as expected for left-handed flux ropes, e.g.,][]{green2007,lynch2009}.

\subsection{Inner Heliosphere} \label{subsec:helio}

\begin{figure}[ht]
\centering
\includegraphics[width=.99\linewidth]{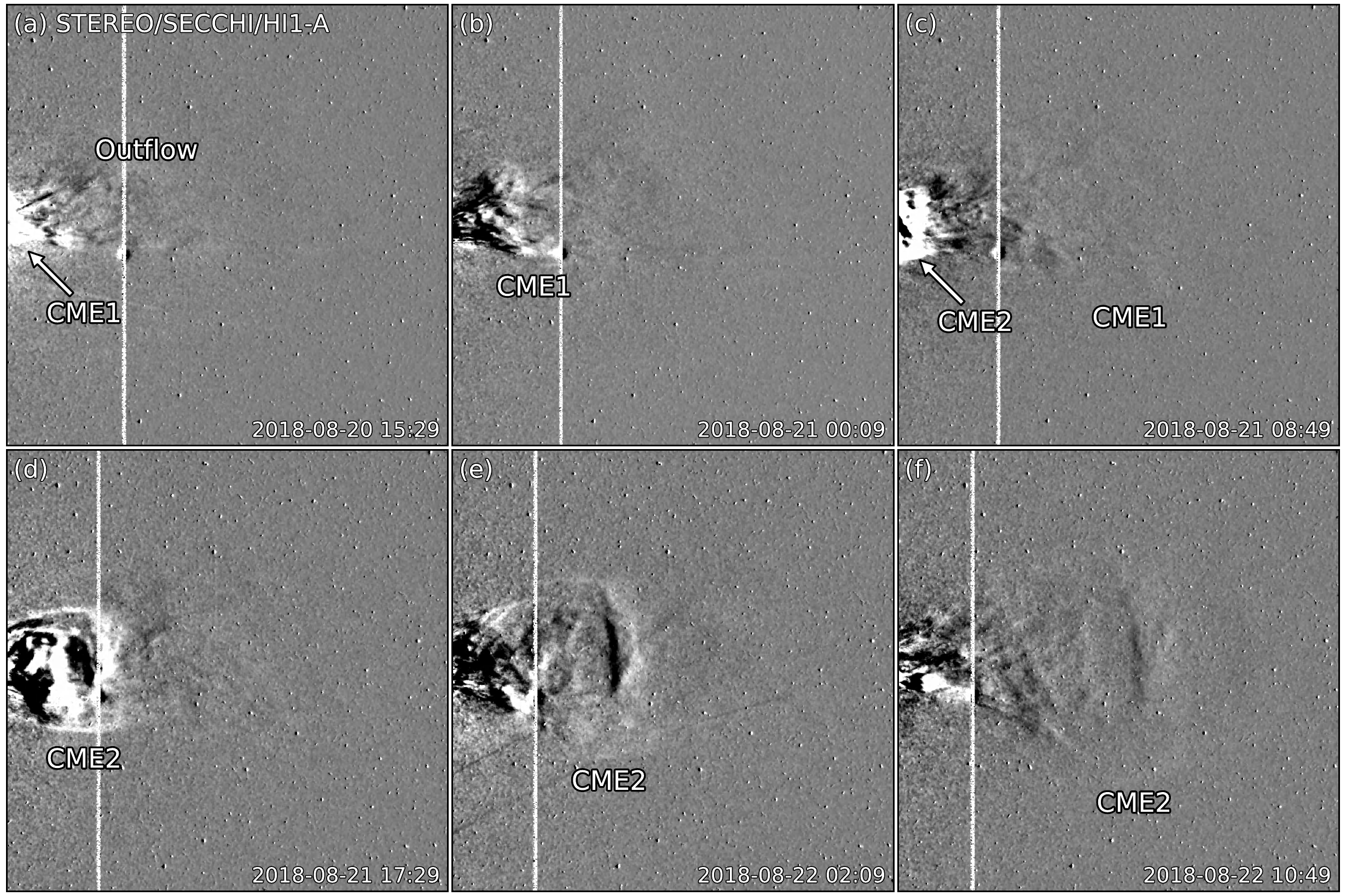}
\caption{Overview of the 2018~August~20 eruptions from remote-sensing inner heliospheric imagery (shown here in running difference with $\Delta t = 40$~min). The locations of the preceding outflow, CME1, and CME2 are indicated in each panel, where appropriate. The vertical band visible throughout the image sequence is caused by Mercury.}
\label{fig:hi}
\end{figure}

After leaving the COR2-A field of view, the eruptions under analysis were observed by the Heliospheric Imager \citep[HI;][]{eyles2009} cameras onboard STEREO-A. Figure~\ref{fig:hi} shows an overview of the observations of CME1 and CME2 through the STEREO/SECCHI/HI1-A camera, whilst the complete set of images is provided in Movie~S4.

CME1 first appeared in the HI1-A field of view at 15:29~UT on 20 August, whilst CME2 emerged at 06:09~UT on 21 August. CME1 maintained a rather irregular morphology, similar to that observed in COR2-A imagery (see Figure~\ref{fig:corona}(a--c) and Movie~S2). CME2, on the other hand, displayed a significantly flatter front than the one featured in COR2-A observations (see Figure~\ref{fig:corona}(e--g) and Movie~S2), likely due to pancaking, i.e.\ flattening of the CME cross-section as it propagates through a latitudinally-structured ambient wind \citep[e.g.,][]{owens2006b,riley2004}. Furthermore, the front of CME1 appeared well ahead of CME2 throughout the sequence of frames in which both eruptions are visible, hence we do not observe signatures of CME--CME interaction in HI1-A imagery.

\section{Heliospheric Context} \label{sec:enlil}

In order to explore the propagation of CME1 and CME2 within their corresponding inner heliospheric context, we employ the Wang--Sheeley--Arge \citep[WSA;][]{arge2004} coronal model coupled with the Enlil \citep{odstrcil2003} heliospheric model, hereafter WSA--Enlil. WSA is used to generate the solar wind background from synoptic magnetogram maps \citep[in this case, from the Global Oscillation Network Group or GONG;][]{harvey1996}, which Enlil takes as input at its inner boundary (set at $21.5\,R_{\odot}$ or 0.1~AU). Enlil then models the heliospheric conditions outwards to an outer boundary (in this case set at 2~AU) by solving the magnetohydrodynamic (MHD) equations. CMEs are inserted in WSA--Enlil at the heliospheric inner boundary as hydrodynamic structures, i.e.\ lacking an internal magnetic field. We derive the input parameters for CME1 and CME2 (shown in Table~\ref{tab:cmeinput}) entirely from the GCS reconstructions outlined in Section~\ref{subsec:corona} and Figure~\ref{fig:corona}. The latitudes ($\theta$), longitudes ($\phi$), and tilts ($\gamma$) are taken directly from their values at the last GCS reconstructions before the CMEs left the COR2-A field of view (at the times shown in the rightmost column of Figure~\ref{fig:corona}). Both CMEs are inserted with an elliptical cross-section, and their semi-major ($\psi_{1}$) and semi-minor ($\psi_{2}$) angular extents are obtained by ``cutting a slice'' out of the GCS shell \citep[see][for details]{thernisien2011}. Finally, the CME insertion speeds ($V_{0}$) are calculated from the CME apex height at the time of the last GCS reconstruction (again, shown in the rightmost column of Figure~\ref{fig:corona}) and the height obtained at the reconstruction performed one hour earlier, and the insertion times ($t_{0}$) are estimated by propagating the ``final'' CME apex until $21.5\,R_{\odot}$ assuming constant speed $V_{0}$.

\begin{table}[!ht]
\centering
\caption{CME Input Parameters for the WSA--Enlil Simulation Run.}
\label{tab:cmeinput}
\centering
\begin{tabular*}{.8\textwidth}{c @{\extracolsep{\fill}} c c c c c c c}
\toprule
CME & $t_{0}$ & $\theta$  & $\phi$ & $\gamma$ & $\psi_{1}$ & $\psi_{2}$ & $V_{0}$ \\
\# & [UT] & [$^{\circ}$]  & [$^{\circ}$] & [$^{\circ}$] & [$^{\circ}$] & [$^{\circ}$] & [km$\cdot$s$^{-1}$]\\
\midrule
1 & 2018-08-20T18:16  & $12$ & $2$ & $10$ & $21$ & $12$ & $483$\\
2 & 2018-08-21T10:56 & $5$ & $10$ & $9$ & $30$ & $16$ & $290$\\
\bottomrule
\end{tabular*}
\begin{tablenotes}
\textit{Note.} The table shows, from left to right: CME number, time ($t_{0}$) of insertion of the CME at the Enlil inner boundary of $21.5\,R_{\odot}$ or 0.1~AU, latitude ($\theta$) and longitude ($\phi$) of the CME apex in Stonyhurst coordinates, tilt ($\gamma$) of the CME axis with respect to the solar equator (defined positive for counterclockwise rotations), semi-major ($\psi_{1}$) and semi-minor ($\psi_{2}$) axes of the CME cross-section, and CME speed at $21.5\,R_{\odot}$ ($V_{0}$).
\end{tablenotes}  
\end{table}

\begin{figure}[t]
\centering
\includegraphics[width=.9\linewidth]{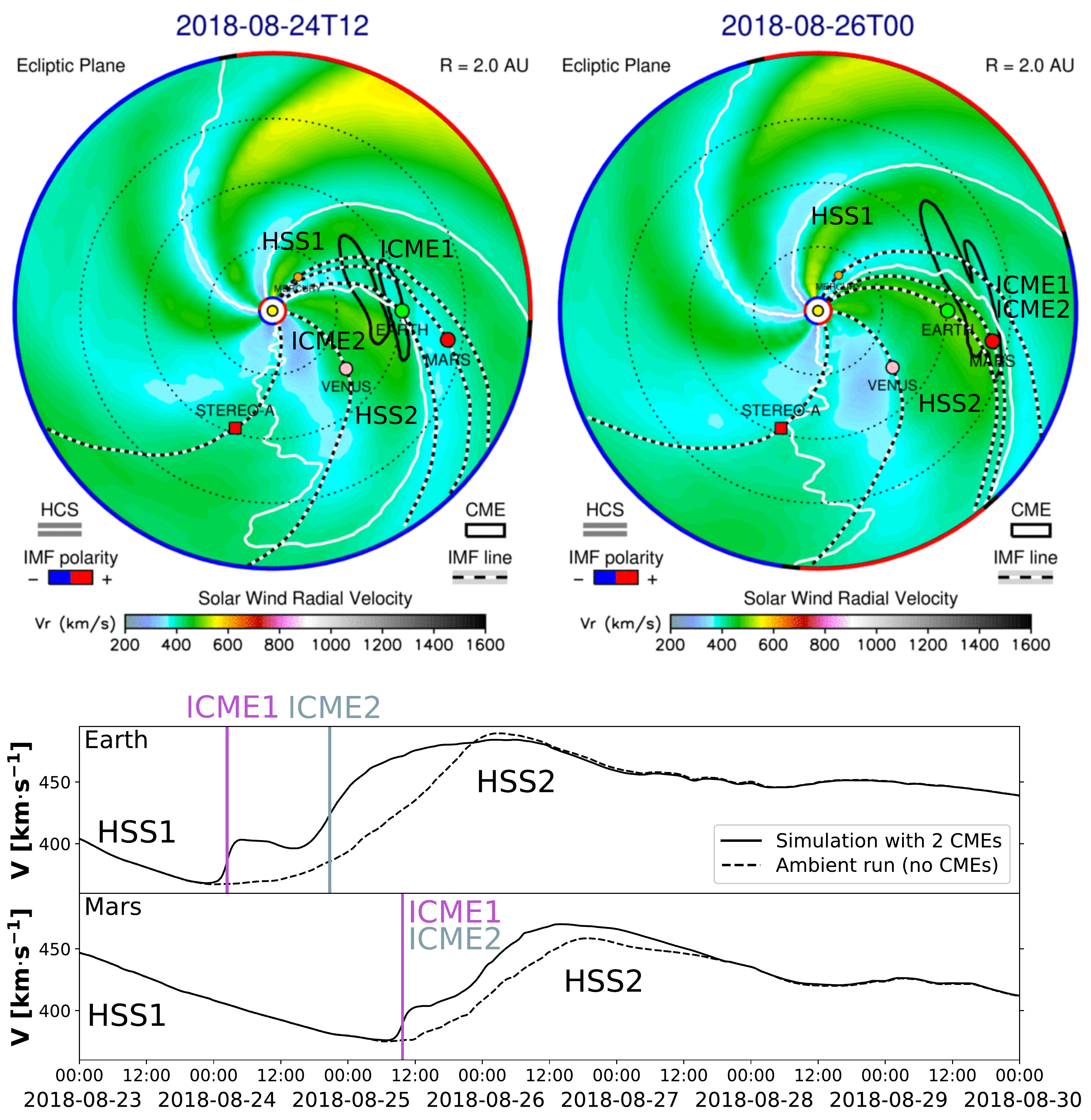}
\caption{Overview of the WSA--Enlil simulation results at Earth and Mars. Top: Snapshots of the simulation results for the radial speed $V_{r}$ on the ecliptic plane around the arrival times at (left) Earth and (right) Mars. Bottom: Results for the solar wind speed $V$ at Earth and Mars.}
\label{fig:enlil}
\end{figure}

An overview of the WSA--Enlil simulation results is shown in Figure~\ref{fig:enlil} and a full animation is shown in Movie~S5. The top panels show snapshots of the solar wind radial speed ($V_{r}$) on the ecliptic plane, from which it is evident that both CME1 and CME2 are ``sandwiched'' between two fast streams marked as `HSS1' and `HSS2', which we attribute to CH1 and CH2, respectively (see Section~\ref{subsec:disc} and Figure~\ref{fig:sun}). `ICME1' and `ICME2' refer to the interplanetary CMEs \citep[ICMEs; e.g.,][]{kilpua2017b} counterparts of CME1 and CME2, respectively. Note that here we refer to ICME as the interplanetary structure as a whole, often composed of a shock, a sheath, and an ejecta. Hence, the ICME arrival time considered here corresponds to the interplanetary shock arrival. By complementing the overall simulation results on the ecliptic plane with the synthetic solar wind speed ($V$) measurements at Earth and Mars shown in the bottom panels of Figure~\ref{fig:enlil}, it is clear that ICME1 and ICME2 would be expected to impact Earth as successive, largely separate structures, whilst the two would have merged by the time they reach Mars. \editone{In the context of the events considered here, there are two possible reasons as to why the slower CME2 would catch up with the initially faster CME1: (1) Due to solar wind preconditioning, resulting in a rarefied ambient medium after the passage of CME1 and, thus, diminished drag \citep[e.g.,][]{temmer2015}, as well as (2) due to the presence of HSS2 trailing CME2, resulting in less deceleration or even acceleration \citep[e.g.,][]{winslow2021b}.} We approximately estimate ICME1 to hit Earth on 2018-08-24T02:25, ICME2 to hit Earth on 2018-08-24T20:45, and the combined ICME1+ICME2 structure to hit Mars on 2018-08-25T09:45.

These evaluations and estimates will be used as support to interpret the in-situ measurements at both Earth and Mars, presented in Section~\ref{sec:insitu}. \editone{Nevertheless, there are a couple of caveats to consider when analysing these results. For example, the lack of an internal magnetic field in the modelled CMEs inevitably results in unrealistic CME--CME interaction outcomes, since the resulting structure simply corresponds to the superposition of two hydrodynamic pulses. In fact, the ICME ejecta boundaries identified in the simulation (black contours in Figure~\ref{fig:enlil}) are estimated in the model via the so-called `cloud tracer' parameter, which tracks the injected mass based on the density enhancement with respect to the background solar wind, i.e.\ they can be considered merely an approximation of the spatial extent of a fully-magnetised ejecta.}


\section{In-Situ Observations} \label{sec:insitu}

In this section, we show and analyse in-situ observations following the solar events described in Section~\ref{sec:remotesensing} at both Earth (Section~\ref{subsec:earth}) and Mars (Section~\ref{subsec:mars}). Then, in Section~\ref{sec:discussion}, these measurements will be synthesised and discussed in relation to the heliospheric context presented in Section~\ref{sec:enlil}.

\subsection{Measurements at Earth} \label{subsec:earth}

\begin{figure}[p]
\centering
\includegraphics[width=.99\linewidth]{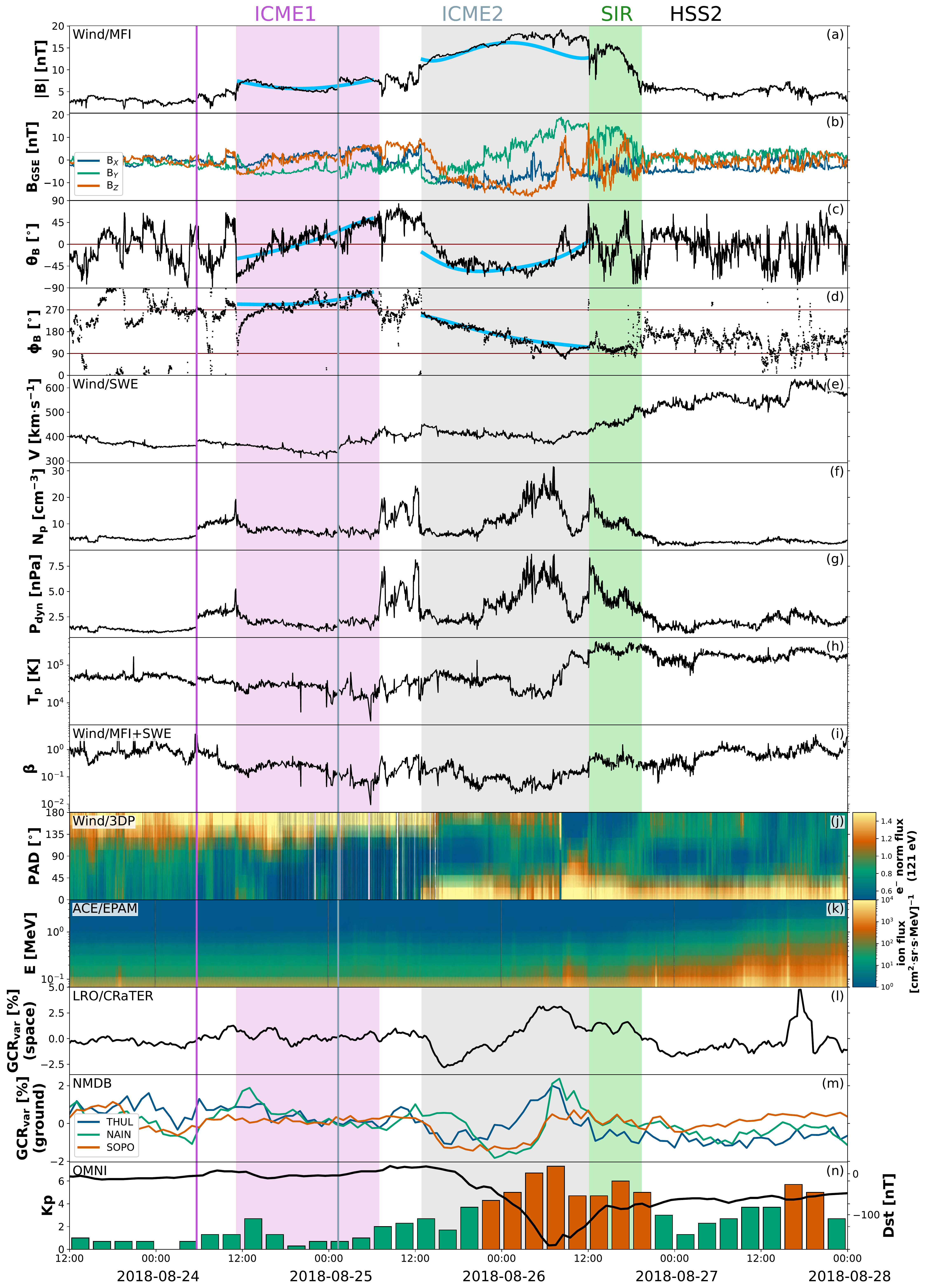}
\caption{In-situ measurements at Earth, showing (a) magnetic field magnitude, (b) magnetic field components in GSE coordinates, (c) $\theta$ and (d) $\phi$ components of the magnetic field, solar wind (e) speed, (f) density, (g) dynamic pressure, and (h) temperature, (i) plasma $\beta$, (j) electron pitch angle distribution, (k) ion intensities, galactic cosmic ray variation (l) in space and (m) on ground, and (n) Kp and Dst indices, quantifying geoeffectiveness. The ejecta intervals show flux rope fitting results with the Elliptic-Cylindrical model superposed on the magnetic field data.}
\label{fig:earth}
\end{figure}

In-situ measurements at Earth are shown in Figure~\ref{fig:earth}. They include: magnetic field data from the Magnetic Field Investigation \citep[MFI;][]{lepping1995}, plasma data from the Solar Wind Experiment \citep[SWE;][]{ogilvie1995}, and electron pitch angle distribution (PAD) data from the Three-Dimensional Plasma and Energetic Particle Investigation \citep[3DP;][]{lin1995} instruments onboard the Wind \citep{ogilvie1997} spacecraft at the Sun--Earth L1 point; suprathermal and energetic ion data from the Electron, Proton, and Alpha Monitor \citep[EPAM;][]{gold1998} onboard the Advanced Composition Explorer \citep[ACE;][]{stone1998} also at the Sun--Earth L1 point; space-based galactic cosmic ray (GCR) estimates from the Cosmic Ray Telescope for the Effects of Radiation \citep[CRaTER;][]{spence2010} onboard the Lunar Reconnaissance Orbiter \citep[LRO;][]{chin2007} orbiting Luna; ground-based GCR estimates from the Thule, Nain, and South Pole stations part of the Neutron Monitor Database \citep[NMDB;][]{mavromichalaki2011}; as well as Kp index from the National Geophysical Data Center (NGDC) and Dst index from the World Data Center (WDC) for Geomagnetism, Kyoto.

The sequence of events observed at Earth commenced with a weak \editone{(i.e., characterised by small downstream-to-upstream ratios in speed, magnetic field magnitude, and plasma density)} interplanetary shock (solid pink line in Figure~\ref{fig:earth}) followed by a weak flux-rope-like ejecta (shaded pink region in Figure~\ref{fig:earth}), which we attribute to ICME1, i.e.\ the interplanetary counterpart of CME1. We fit the ejecta with the Elliptic-Cylindrical \citep[EC;][]{nieveschinchilla2018b} analytical flux rope model (shown in blue superposed on the Wind/MFI data in Figure~\ref{fig:earth}), resulting in a moderately inclined structure with axis orientation ($\Theta$, $\Phi$) = ($36^{\circ}$, $230^{\circ}$) in Geocentric Solar Ecliptic (GSE) coordinates and right-handed chirality, which roughly corresponds to a SWN-type flux rope. About two-thirds into the ICME1 flux rope, a second weak interplanetary shock (solid grey line in Figure~\ref{fig:earth}) was detected, displaying classic shock-in-ejecta signatures \citep[e.g.,][]{lugaz2015a}: Proton temperature and plasma beta maintained lower-than-ambient values \citep[typical ICME ejecta indicators; e.g.,][]{zurbuchen2006} and all the magnetic field components did not display changes in the clock angle after the shock passage. We attribute this shock to ICME2, i.e.\ the interplanetary counterpart of CME2, suggesting that the two eruptions were at the initial stages of their CME--CME interaction process at the time they travelled past Earth. We note no significant particle enhancements \editone{in ACE/EPAM data}, no Forbush decreases \citep[e.g.,][]{forbush1937} in the GCR intensity, and no \editone{major} geomagnetic effects associated with ICME1 \editone{in terms of the Kp and Dst indices}. \editone{Nevertheless, although the relatively low values of speed, dynamic pressure, and magnetic field did not result in a geomagnetic storm, some substorm activity was observed, with the AE index peaking at ${\gtrsim}700$~nT (not shown), likely in response to the period of weakly negative $B_{Z}$ in the leading portion of the ICME ejecta.}

After a brief period of high density and dynamic pressure measured by Wind/SWE \editone{as well as a rather weak but turbulent magnetic field measured by Wind/MFI}, \editone{which has been interpreted as a signature of interaction between two successive CMEs \citep{lugaz2005b, lugaz2017a, wang2003}} and evidently part of the ICME2 sheath, a clear flux-rope ejecta (shaded grey region in Figure~\ref{fig:earth}) was detected. Fitting results with the EC model (shown in blue below the Wind/MFI data in Figure~\ref{fig:earth}) yield a left-handed flux rope with a rather high inclination and central axis orientation ($\Theta$, $\Phi$) = ($-70^{\circ}$, $145^{\circ}$) in GSE coordinates, corresponding roughly to a WSE-type rope. Contrary to ICME1, ICME2 featured clear counterstreaming electron signatured in Wind/3DP data and was accompanied by a weak (${\sim}2$\% drop) Forbush decrease detected both in space by LRO and on ground by different neutron monitors, albeit with slightly different profiles---the LRO/CRaTER data are closest to the profile measured on ground at the Thule station. Most strikingly, ICME2 was particularly geoeffective, with a maximum Kp of 7+ (corresponding to a G3 storm, see \url{https://www.swpc.noaa.gov/noaa-scales-explanation}) and a minimum Dst of $-175$~nT \citep[well exceeding the usual threshold for ``strong'' storms of Dst$_\mathrm{min} \leq -100$~nT; e.g.,][]{zhang2007}, \editone{while the AE index peaked at ${\gtrsim}2200$~nT (not shown)}. These factors make ICME2 the driver of a so-called ``problem geomagnetic storm'' \citep{nitta2021}, not because its source was stealthy or elusive \citep[see][for a discussion on CMEs without appreciable low-coronal signatures on the Sun]{nitta2017}, but because its effects at Earth were largely unexpected given the slow and ``unimpressive'' nature of CME2 at the Sun. In fact, the solar wind speed remained around values of ${\sim}400$~km$\cdot$s$^{-1}$ throughout the ICME passage, indicating that the storm was mostly driven by the sustained southward $B_{Z}$ magnetic field component within the ejecta, which peaked at $-16$~nT. \editone{Other factors} contributing to the intensification of the geomagnetic storm may have been \editone{the interaction with the preceding ICME1, leading to compression at the front of the ejecta, as well as} the presence of a nearby CH (in this case CH2, see Figure~\ref{fig:sun}) back at the Sun, resulting in CME--HSS interaction in interplanetary space \editone{and compression in the trailing part of the ejecta} \citep[e.g.,][]{nitta2021}.

Indeed, the ICME2 ejecta was immediately followed by a small SIR (shaded green region in Figure~\ref{fig:earth}) and related HSS, most likely due to CH2 and thus indicated in Figure~\ref{fig:earth} as HSS2. That ICME2 and HSS2 were in the process of interacting is evident by the increasing solar wind speed and temperature detected by Wind/SWE, as well as the particle enhancement at suprathermal energies measured by ACE/EPAM in the trailing part of the ejecta. By the time Earth was impacted by HSS2, the intense geomagnetic storm had essentially waned, with the Kp index briefly reaching again values up to $6{-}$ almost a full day later. \editone{Nevertheless, a sequence of substorms persisted through the passage of the SIR and the initial portion of the following HSS2, with the AE index peaking between ${\sim}1000$ and ${\sim}1500$~nT for each event (not shown).}

\subsection{Measurements at Mars} \label{subsec:mars}

\begin{figure}[p]
\centering
\includegraphics[width=.99\linewidth]{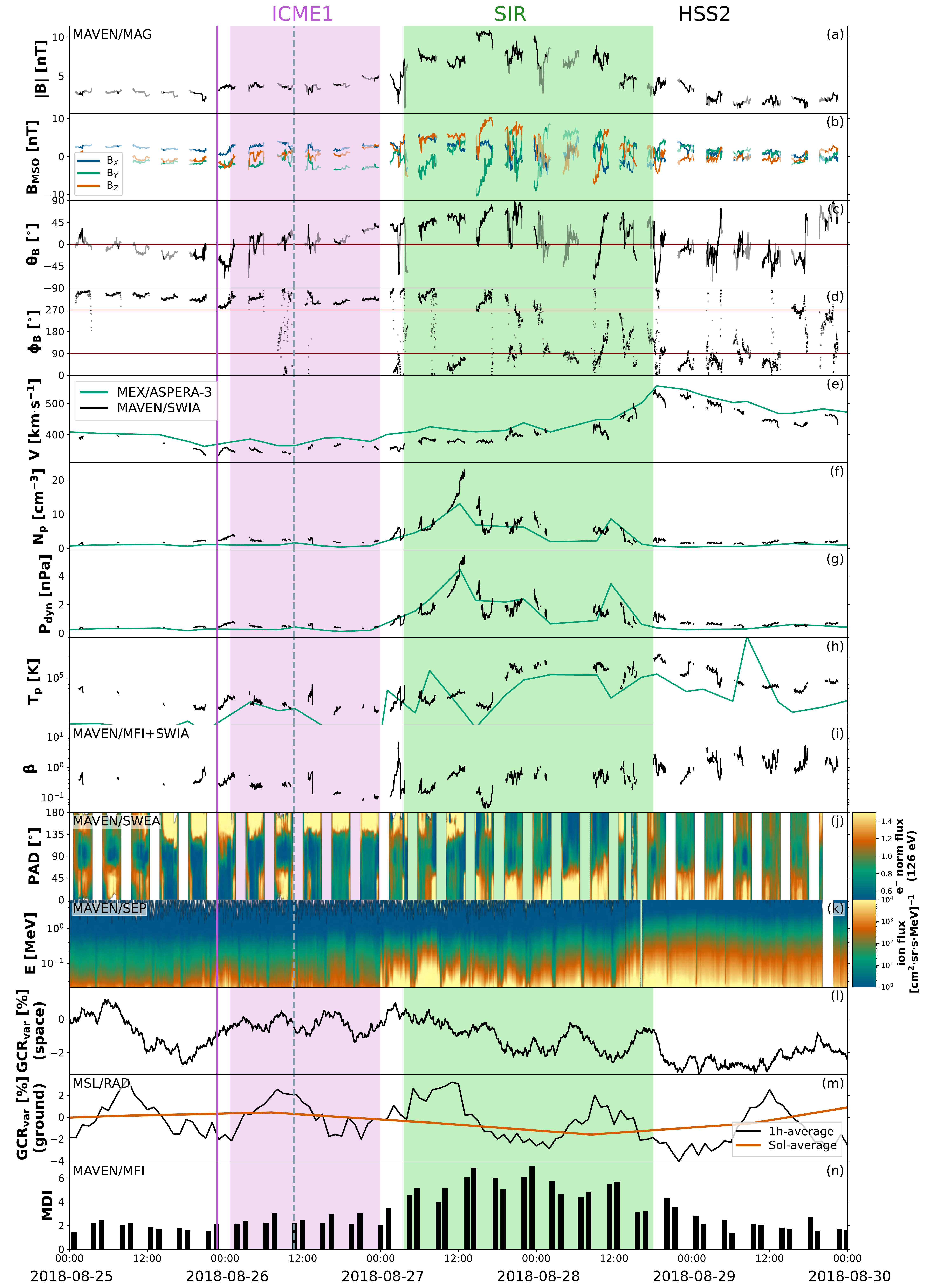}
\caption{In-situ measurements at Mars, showing (a) magnetic field magnitude, (b) magnetic field components in MSO coordinates, (c) $\theta$ and (d) $\phi$ components of the magnetic field, solar wind (e) speed, (f) density, (g) dynamic pressure, and (h) temperature, (i) plasma $\beta$, (j) electron pitch angle distribution, (k) ion intensities, galactic cosmic ray variation (l) in space and (m) on ground, and (n) MDI index, quantifying arieffectiveness. The lighter portions of magnetic field data are sampled in the Martian foreshock rather than in the solar wind.}
\label{fig:mars}
\end{figure}

In-situ measurements at Mars are shown in Figure~\ref{fig:mars}. They include: magnetic field data from the Magnetometer \citep[MAG;][]{connerney2015}, plasma data from the Solar Wind Ion Analyzer \citep[SWIA;][]{halekas2015}, electron PAD data from the Solar Wind Electron Analyzer \citep[SWEA;][]{mitchell2016}, and suprathermal/energetic ion data as well as space-based GCR estimates from the Solar Energetic Particle \citep[SEP;][]{larson2015} instruments onboard the Mars Atmosphere and Volatile Evolution \citep[MAVEN;][]{jakosky2015b} spacecraft orbiting Mars; plasma data from the Analyzer of Space Plasmas and Energetic Atoms \citep[ASPERA-3;][]{barabash2006} onboard Mars Express \citep[MEX;][]{chicarro2004} also orbiting Mars; ground-based GCR estimates from the Radiation Assessment Detector \citep[RAD;][]{hassler2012} onboard the Curiosity rover part of the Mars Science Laboratory \citep[MSL;][]{grotzinger2012}; and Magnetospheric Disturbance Index \citep[MDI;][]{gruesbeck2021} data. MDI takes MAVEN/MAG measurements as input and is used to estimate the level of the disturbance to the Martian system, or arieffectiveness.

The sequence of events at Mars commenced again with a \editone{possible} weak interplanetary shock (solid pink line in Figure~\ref{fig:mars}), identified mainly via a peak in ion fluxes at suprathermal energies detected by MAVEN/SEP because of the frequent gaps in magnetic field and plasma data. In fact, these data gaps are due to the MAVEN and MEX orbits, resulting in only a portion of their path around Mars sampling the upstream solar wind. The MEX/ASPERA-3 measurements displayed in Figure~\ref{fig:mars} consist of averages along the spacecraft's orbit, and the MAVEN/MAG and MAVEN/SWIA data show solar wind periods identified according to the algorithm of \citet{halekas2017}. The MAVEN/MAG measurements shown in lighter colours represent data that are not part of the ``undisturbed'' solar wind but are still outside the Martian bow shock, i.e.\ they are collected in the foreshock region. The \editone{possible} interplanetary shock is followed by a weak ejecta (shaded pink region in Figure~\ref{fig:mars}), with directions of the magnetic field vectors closely resembling those observed at Earth for ICME1 (see Figure~\ref{fig:earth}). The $\theta_{B}$ component rotates from south to north and the $\phi_{B}$ one points to the west in Mars Solar Orbital (MSO) coordinates (i.e., the GSE equivalent for Mars), indicating a SWN-type flux rope. Considering the ICME1 shock speed at Earth (${\sim}385$~km$\cdot$s$^{-1}$) and the Earth--Mars radial distance at the time of ${\sim}0.4$~AU, the structure would have taken ${\sim}43$~hours to propagate from one planet to the next, in agreement with the timing of the observed \editone{candidate} shock at Mars. Hence, we attribute this structure to ICME1. Approximately in the middle of the ejecta, we identify a possible second interplanetary shock (dashed grey line in Figure~\ref{fig:mars}), again because of a small enhancement in the MAVEN/SEP spectrum and because the solar wind speed measured by MAVEN/SWIA displays a small increase afterwards. This is likely the ICME2 shock, still travelling through the ICME1 ejecta---again, the timing (${\sim}44$~hours for a shock moving at ${\sim}380$~km$\cdot$s$^{-1}$ across ${\sim}0.4$~AU) is consistent between the observations at Earth and Mars. ICME1 was characterised by some periods of bidirectional electrons measured by MAVEN/SWEA, especially during the first half of the ejecta, no Forbush decreases in space nor on ground, and no significant disturbances to the Martian system.

After the passage of the weak ICME1, a period of clear enhanced magnetic field magnitude followed (shaded green region in Figure~\ref{fig:mars}). However, this structure displays classic signatures of a SIR rather than of an ICME \citep[e.g.,][]{kataoka2006}, including the absence of an organised internal magnetic field throughout its extent, and an increase in solar wind speed and temperature as well as a decrease in density at the stream interface. This suggests that the ejecta of ICME2, which was rather prominent at Earth (see Figure~\ref{fig:earth}), largely missed Mars. It is unclear whether minor portions of the ejecta interacted and/or merged with the following SIR at Mars's heliolongitude, especially given the numerous data gaps throughout the structure. \editone{The plasma profiles measured by MAVEN/SWIA in the first half of the SIR resemble those observed at Earth in the ICME2 ejecta and following SIR (see Figure~\ref{fig:earth}), but the absence of bidirectional electrons in MAVEN/SWEA data as well as the disorganised magnetic field components seen by MAVEN/MAG suggest that, if material from CME2 was indeed detected at Mars, the ejecta had lost its coherence and flux rope structure (at least locally) due to interaction with the structured solar wind.} The SIR was accompanied by enhanced ion fluxes measured by MAVEN/SEP at sub-MeV energies \citep[see][for a review on SIR-associated energetic particle observations]{richardson2004b}, as well as by a weak (${\sim}2$\% drop) Forbush decrease measured both in space by MAVEN/SEP and on ground by MSL/RAD \citep[see][for an overview of space- and ground-based observations of CME- and SIR-driven Forbush decreases at Mars]{guo2018a}. We note that no significant variations were detected in the MEX/ASPERA-3 Ion Mass Analyser (IMA) background counts (not shown), which can be used as proxies for GCR intensity \citep[see][for a description of the data set]{futaana2022}. Additionally, the SIR was rather arieffective according to MDI values, peaking at ${\sim}7$---we note that \citet{gruesbeck2021} reported a peak MDI of ${\sim}4$ for the 8 March 2015 ICME studied by \citet{jakosky2015a} and of ${\sim}10$ for the 13 September 2017 ICME analysed by \citet{lee2018}. Finally, as expected, the SIR was ultimately followed by HSS2.


\section{Discussion} \label{sec:discussion}

The eruptive events that took place during the second half of August 2018 were typical of solar minimum conditions: They did not originate from active regions nor did they display an ``explosive'' nature, they were slow, and they were rather ``isolated'' (i.e., there was no other activity elsewhere on the Sun during the same period). Nevertheless, such a seemingly simple picture resulted in a number of unexpected outcomes, including a major problem storm at Earth and a missed CME impact at Mars, despite the small (${<}10^{\circ}$) longitudinal separation between the two planets. Here, we synthesise the solar and heliospheric observations discussed in the previous sections together with the WSA--Enlil modelling results, and elaborate on the possible evolutionary scenario that is consistent with measurements at both planets.

First of all, we note that the CME arrival times estimated by WSA--Enlil (see Figure~\ref{fig:enlil}) are between ${\sim}4$~hours (ICME1 at Earth) to ${\sim}12$~hours (ICME1 at Mars) too early compared to observations. This is likely due to HSS2 impacting Earth approximately 1~day too early and Mars approximately 2~days too early in the simulation, resulting in a diminished deceleration of the CMEs ahead due to solar wind drag. Nevertheless, \citet{gressl2014} found that predicted HSS arrival times in MHD models have uncertainties of the order of about one day, suggesting that the period under analysis was not particularly challenging in terms of characterising the solar wind background, since it was affected by typical errors. Arrival times aside, the sequence of events at Earth was well reproduced in the simulation (see Movie~S5 for a more complete view of the modelling results): ICME2 closely followed ICME1 (and the eruptions had just started to interact by the time they crossed Earth) and was immediately trailed by HSS2. What was impossible to forecast from the WSA--Enlil simulation was the impressive magnitude of the geomagnetic storm driven by ICME2.

\begin{figure}[th!]
\centering
\includegraphics[width=.99\linewidth]{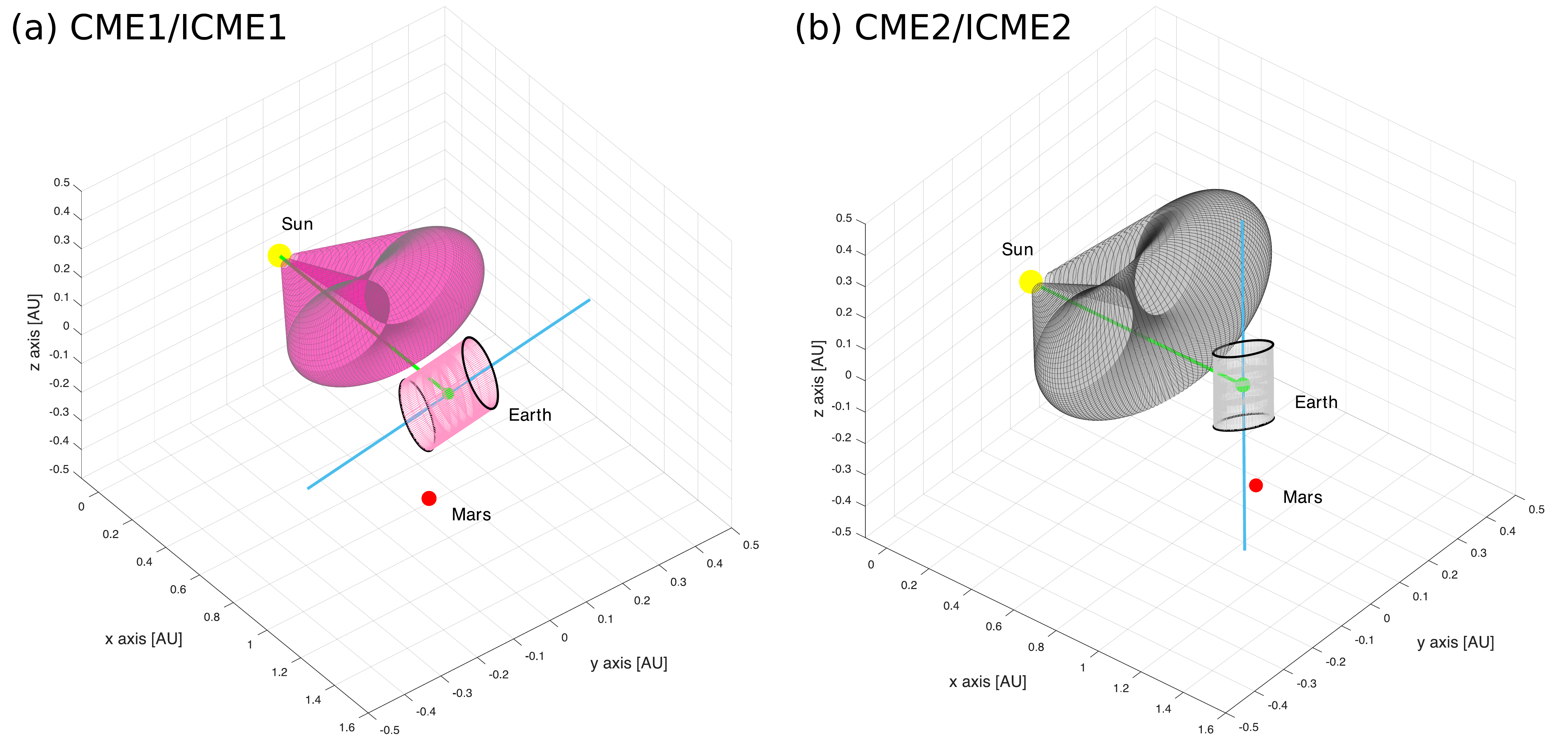}
\caption{Comparison of the GCS-derived CME orientation in the solar corona and EC-derived ICME ejecta orientation at Earth for (a) CME1/ICME1, in pink, and (b) CME2/ICME2, in grey. The croissant-like features show the CME morphology obtained via the GCS reconstructions shown in Figure~\ref{fig:corona}, whilst the cylinders represent the corresponding ICME structures retrieved via the EC fits displayed in Figure~\ref{fig:earth}.}
\label{fig:orientations}
\end{figure}

In fact, the factors that led to the unexpected geoeffectiveness of ICME2 at Earth may also explain why the same structure missed Mars. This is illustrated in Figure~\ref{fig:orientations}, which shows a comparison of the CME flux rope orientations retrieved from solar observations (i.e., the GCS reconstructions in Figure~\ref{fig:corona}) and in-situ measurements at Earth (i.e., the EC fits in Figure~\ref{fig:earth}). It is evident that CME1/ICME1 maintained a low inclination to the equatorial plane throughout its heliospheric evolution, resulting in a flux rope type that was estimated to be SWN in the solar corona, at Earth, and later at Mars. CME2, on the other hand, left the Sun with a WSE configuration but was later observed in the solar corona to have assumed a lowly-inclined orientation (possibly NWS, see especially its morphology in SOHO/LASCO data in Figure~\ref{fig:corona}(m--p)). At Earth, however, the retrieved flux rope type of ICME2 was again WSE, leading to the strong southward field that mainly drove the observed geomagnetic storm. We speculate that interaction with the following HSS2 resulted in both CME rotation towards a high-inclination configuration and CME deflection towards western heliolongitudes. The latter effect is marginally visible in the WSA--Enlil simulation (see Movie~S5), whilst the former could not be reproduced with the current set up since the two CMEs have been modelled without an internal magnetic field, leading to unrealistic CME--CME and CME--HSS interaction processes. The rotation and deflection of the CME2 flux rope ejecta in interplanetary space may account for the fact that only the ICME2 shock seems to have impacted Mars. \editone{Of course, these considerations imply that CME2 was indeed characterised by a low inclination with respect to the solar equator throughout the solar corona---since the GCS reconstructions were performed using the only two available viewpoints of Earth (halo event) and STEREO-A (limb event), the resulting morphology may have been under-constrained and thus associated with rather large uncertainties.}

Another interesting result is that, whilst Mars \editone{was not impacted by} the ICME2 ejecta, it was still affected by significant large-scale disturbances during the period under study, namely due to the SIR ahead of HSS2. At Earth, the SIR was not particularly geoeffective, suggested by the fact that the Dst index did not display a second dip after the passage of ICME2 (although it does appear to have interrupted the recovery of the ongoing storm)---it is worth noting, however, that the presence of ICME2 at Earth likely impeded the formation of a proper slow--fast stream interaction region, and that what we identified as a small SIR displays characteristics of a CME--HSS interaction region. At Mars, the SIR passage lasted for over a day and was fundamentally responsible for the observed arieffectiveness. \editone{We remark, nevertheless, that the plasma signatures in the first half of the SIR at Mars displayed similarities with those observed in the ICME2 ejecta at Earth, possibly indicating that some material from CME2 may have become part of the SIR due to interaction with the structured solar wind.} Whilst it is not known whether similar effects would have been experienced at Mars had the ICME2 ejecta impacted the planet \editone{as a well-defined flux rope structure}, we can speculate that, during solar minimum conditions, when CMEs are on average less energetic, we may generally expect higher arieffectiveness from SIRs than from ICMEs \citep[contrarily to Earth, where the most geoeffective structures are associated with CMEs across the whole solar cycle; e.g.,][]{richardson2001b}. In fact, whilst CMEs tend to expand (and thus experience a decrease in speed, density, and dynamic pressure) until ${\sim}10$--15~AU, where they reach pressure balance with the ambient wind \citep[e.g.,][]{vonsteiger2006}, SIRs are known to increase the compression of the fast stream against the slow stream with heliocentric distance until a few AUs \citep[e.g.,][]{forsyth1999}---for example, \citet{jian2008a} found a significant increase in SIR-associated shock detections from Venus to Earth, and \citet{geyer2021} found that the occurrence of SIR-driven fast-forward shocks is three times higher at Mars than at Earth. Additionally, \citet{edberg2010} found that most solar wind pressure pulse events---which lead to a 2.5-fold increase in atmospheric escape---correspond to SIR drivers, whilst only a few are driven by CMEs.


\section{Concluding Remarks} \label{sec:conclusion}

In this work, we have analysed the eruption, evolution, and impact at Earth and Mars of a series of solar transients during the second half of August 2018, i.e.\ close to the activity minimum between solar cycles 24--25. In particular, we have tracked the eruption (on 20 August 2018) and propagation of two CMEs (CME1 and CME2) throughout their journey in the inner heliosphere as they were followed by a fast solar wind stream (HSS2). We found that both planets experienced space weather effects, but due to different drivers: at Earth, an ICME ejecta (ICME2) with a strong southward component of its magnetic field was the main driver of a major geomagnetic storm, whilst at Mars, the \editone{main} structure responsible for the observed arieffectiveness was a SIR. The first event in the sequence of transients, i.e.\ a preceding, smaller ICME (ICME1), was detected at both planets but had no significant effects on their space environments.

We also compared in-situ measurements at Earth and Mars with simulation results using the MHD WSA--Enlil model. First of all, we noted that the observed geomagnetic storm could not have been predicted with the used set up, since the CMEs were modelled without an internal magnetic field. Heliospheric models that include a magnetised CME are expected to provide more realistic insights into the dynamics of CME--CME and CME--HSS interaction \citep[e.g.,][]{asvestari2022,scolini2020}. The hydrodynamic nature of the modelled CMEs may also explain why CME2 was predicted to merge with CME1 beyond 1~AU and ultimately impact Mars, whilst it likely rotated and drifted towards western heliolongitudes due to interactions with the following HSS2.

At Mars, the SIR-induced effects would have been predicted more successfully with the modelling set up used in this study, although the HSS of interest (HSS2) was simulated to arrive earlier than observed at both planets. \citet{riley2012} found that global MHD models can diverge more or less significantly from observations in capturing the overall structure of the ambient solar wind even for solar minimum periods, when conditions are relatively steady. \citet{jian2011} reported that SIR timing predictions can have temporal offsets of up to two days at 1~AU and up to four days at 5~AU. Large-scale efforts dedicated to benchmarking solar wind models \citep[e.g.,][]{reiss2022} will likely lead to improved predictions of solar wind structures (including SIRs and CMEs) and their effects at different planets. \editone{In fact, a well-constrained and well-reproduced solar wind background is of great importance for simulating solar minimum CMEs, which tend to alter their structure and orientation during propagation largely due to interactions with the steady wind \citep[see, e.g., the May 1997 event;][]{cohen2010,odstrcil2004,titov2008}.}

Finally, we comment on two aspects that may be of interest to martian space weather research. First, although most major geomagnetic storm are associated with CMEs, it has been reported that SIRs can occasionally drive strong responses at Earth \citep{richardson2006b}. Given that SIRs are expected to strengthen beyond 1~AU, an interesting exercise would be to follow one such strongly geoeffective structure from Earth to Mars and evaluate its arieffectiveness---as space weather at Mars is a relatively novel area of research, it is currently unclear which solar wind transients are generally associated with the most severe disturbances. Additionally, the different types of magnetospheres may lead to different solar wind driving conditions at Earth and Mars. Of course, analysis of the solar wind drivers of martian storms would strongly benefit from continuous space weather monitoring, perhaps via a satellite placed at the Sun--Mars L1 point as is the case for Earth \citep[e.g.,][]{sanchezcano2021b}.


\section*{Data Availability Statement}

Solar disc and coronagraph data from SDO, SOHO, and STEREO are openly available at the Virtual Solar Observatory (VSO; \url{https://sdac.virtualsolar.org/}), whilst STEREO/HI Level-2 data can be retrieved from the UK Solar System Data Centre (UKSSDC; \url{https://www.ukssdc.ac.uk}). These data were visualised, processed, and analysed trough SunPy \citep{sunpy2020}, IDL SolarSoft \citep{bentely1998}, and the ESA JHelioviewer software \citep{muller2017}.
Enlil simulation results have been provided by the Community Coordinated Modeling Center at Goddard Space Flight Center through their public Runs on Request system (\url{http://ccmc.gsfc.nasa.gov}). The full simulation results are available at \url{https://ccmc.gsfc.nasa.gov/results/viewrun.php?domain=SH&runnumber=Erika_Palmerio_050922_SH_1} (run id: \texttt{Erika\_Palmerio\_050922\_SH\_1}).
Wind and ACE data are publicly available at NASA's Coordinated Data Analysis Web (CDAWeb) database (\url{https://cdaweb.gsfc.nasa.gov/index.html/}). NMDB data are publicly available at \url{http://www.nmdb.eu}.
LRO, MAVEN, and MSL data can be accessed at the Planetary Plasma Interactions (PPI) Node of NASA’s Planetary Data System (PDS) database (\url{https://pds-ppi.igpp.ucla.edu}). MEX data are openly available at ESA's Planetary Science Archive (PSA) database (\url{https://archives.esac.esa.int/psa}).


\acknowledgments
E.~Palmerio acknowledges the NASA Living With a Star (LWS) Jack Eddy Postdoctoral Fellowship Program, administered by UCAR's Cooperative Programs for the Advancement of Earth System Science (CPAESS) under award no.\ NNX16AK22G, as well as NASA's HTMS Program (grant no.\ 80NSSC20K1274) and NSF's PREEVENTS Program (grant no.\ ICER-1854790).
C.~O.~Lee, J.~R.~Gruesbeck, J.~S.~Halekas, and S.~Xu acknowledge support from the MAVEN mission, which is supported by NASA through the Mars Exploration Program.
C.~O.~Lee, S.~Xu, B.~J.~Lynch, and J.~G.~Luhmann acknowledge NASA HGI (no.\ 80NSSC21K0731) and LWS (no.\ 80NSSC21K1325).
I.~G.~Richardson acknowledges support from NASA LWS Program NNH19ZDA001N-LWS.
The WSA model was developed by C.~N.~Arge (currently at NASA/GSFC), and the Enlil model was developed by D.~Odstrcil (currently at GMU). We thank the model developers and the CCMC staff.
E.~Palmerio, I.~G.~Richardson, N.~V.~Nitta, T.~Mulligan, and B.~J.~Lynch thank the International Space Science Institute (ISSI) for their support of International Team no. 415, ``Understanding the Origins of Problem Geomagnetic Storms'' (\url{https://www.issibern.ch/teams/geomagstorm/}).

\bibliography{bibliography}

\end{document}